\documentclass[12pt]{article}
\usepackage{graphicx,times,achemso,bm,amssymb}

\title{Phase behavior of C18 monoglyceride in hydrophobic solutions}
\author{C.H.~Chen$^{1}$, 
I. Van Damme$^{2}$ 
        and E.M.~Terentjev$^{1}$   
\\ {  } \\
{ \small $^{1}$ Cavendish Laboratory, University of Cambridge}
\\ {\small J J Thomson Avenue, Cambridge CB3 0HE, U.K.}
 \\  {  } \\
{ \small  $^{2}$ R\&D Mars UK, Dundee Rd, Slough SL1 4JX, U.K.}
 }
\begin{document}
\maketitle
\renewcommand{\thefootnote}{\fnsymbol{footnote}}
\noindent
We apply a set of different
techniques to analyze the physical properties and phase transitions
of monoglycerides (MG) in oil. In contrast to many studies of MG in
water or aqueous systems, we find a significant difference
in the phase structure at different concentrations and temperatures.
By adding small quantities of water to our base MG/oil systems we
test the effect of hydration of surfactant head-groups, and its effect
on the phase behavior. The phase diagrams are determined by calorimetry
and their universal features are recorded under different conditions.
Two ordered phases are reported: the inverse lamellar gel phase and
the sub-alpha crystalline gel phase. This sequence is very different from the
structures in MG/water; its most striking feature is the establishing of a
2D densely packed hexagonal order of glycerol heads in the middle of inverse
lamellar bilayers. Rheology was examined through temperature scans to
demonstrate the gelation phenomenon, which starts
from the onset of the lamellar phase during the
cooling/ordering process.

\newpage
\section{Introduction}


Monoglyceride (MG) is a lipid molecule consisting of a single
fatty acid chain linked to a
glycerol head.\cite{Larsson68,Tiddy93} MG variants are distinguished by the
length of carbon chain and here we focus on a particularly common
surfactant labeled C18. Because of the strong emulsifying property,
MG is widely used in personal care products, cosmetics and in food
industry to tailor product properties in a specifically
manner.\cite{Veeman05,Michel05,book_Aljandro,Ian03,Sagalowicz07} A
lot of manufactured oil-based food products such as chocolate,
cakes and creams contain fat (triglyceride) as a structuring material,
which raises obesity issues. In order to reduce the
fat in the oil-based products, one of the strategies is to achieve the
structuring of oil without fat, by exploiting the
liquid crystalline phase of long-chain saturated MG in
hydrophobic solvents.\cite{Marangoni06b} A particular feature is the
use of MG/oil mixtures as healthy substitute for
butter.\cite{Shimoni04a, Shimoni04b} Apart from specific industrial
applications, MG/oil systems are a part of generic materials,
which form a percolating network of structured aggregates to rend the
solvents inside.\cite{Hendricx98} For these reasons the detailed
knowledge of
phase behavior and micro-structures of MG/oil systems are important
in terms of both scientific reasons and the technological
implementations.

MG aqueous systems have been studied systematically for many years.
The liquid-crystal and fully crystalline structures of pure MG was firstly
described by Larsson in 1966 \cite{Larsson66} and was later reviewed
by Small (1986)\cite{book_Small}, Larsson (1994)\cite{Larsson94},
and Krog (2001)\cite{book_Garti}. Its polymorphic behavior is well
known, with the following phase sequence on cooling: isotropic fluid,
lamellar and alpha-crystalline phases.\cite{Larsson67, Larsson90,
Sein02, Hendricx98} When polar lipids are mixed with water, between
the gelation temperature and the Krafft temperature the water has a
strong affinity to the polar glycerol groups and will penetrate the polar
sheets, forming a lamellar liquid crystalline phase which contains the
lamellar ordering with disordered carbon chains.\cite{Larsson67}
Below the Krafft temperature $T_{K}$ the alkyl chains are partially
frozen and a hydrated mesomorphic phase, named the alpha-crystalline gel
($L_\beta$), is formed.\cite{Larsson97} This phase is characterized by a
single X-ray reflection in wide angles, corresponding to a short spacing
of 4.18\AA, which shows hexagonal packing.\cite{book_Garti, Marangoni06b ,book_Small}

The alpha-crystalline phase is metastable and eventually ages into
the anhydrous MG crystal, named the beta-crystalline state (often
referred to as the ``coagel''),
which has a higher melting point and is characterized by a number of
distinct wide-angle
X-ray reflections, with the strongest line corresponding to the spacing
4.5-4.6\AA.\cite{book_Garti, book_Small} A coagel state of pure MG, and equivalently
 in below the Krafft demixing temperature is due to hydrogen bonds
 establishing within head groups in
bilayers, which in turn lead to a further crystallization of
aliphatic tails.\cite{Agt98} On a long time-scale of aging, the D-
and L- isomers of chiral MG gradually separate within crystalline
bilayers, leading to more dense packing and full expulsion of water.
Sedimentation of solid in this phase then takes places.\cite{Agt98,
Mohwald96}

Comparing with aqueous MG systems, the phase behaviors of MG in a
fully hydrophobic solvent is much less studied. In this case some
rheological and storage properties, and the network features have been
reported by Shimoni et al\cite{Shimoni04a, Shimoni04b, Shimoni07},
but due to the absence of the phase diagram and confident structure
description the connection between the molecular arrangement and
macroscopic observations is weak and need systematic analysis. In
fact, many authors in this field continue to assume that the phase diagram of
MG in oil is the same, or similar, as in water, whereas it is clear
that hydrogen-bonding patterns of aggregated glycerol groups would be very
different.\cite{Shimoni04a, Shimoni04b} There are many examples of
the use of ternary MG/water/oil systems; however, the presence of
water will dominate the phase behavior of aggregating
MG.\cite{Marangoni07a, Marangoni06b, Marangoni07c} Some of our
X-ray results reported here are similar to those of
Marangoni et al,\cite{Marangoni06b} but
once again our present work is
different since we deliberately conducted studies with no water
present.

In the case of MG/water, the molecules pack in usual hydrophilic
lamellae, with alkyl chains packing inside the bilayer, which in turn
are surrounded by water, binding to the glycerol heads by hydrogen bonds.
In the lamellar phase the head groups are
in thermal motion, and the aliphatic tails inside the bilayer are
also molten. It is a well-studied liquid-crystalline system. In particular,
considering the hydrophobic chains of each monolayers as an extended polymer
brush, it is clear that normal pressure on the bilayer
would not change the structure of chains in such a brush.\cite{Klein05}
In contrast, in the hydrophobic environment, the
inverse lamellar bilayer structure will form, with aliphatic tails on the
outside the bilayer, also in the molten extended-brush conformation. However,
the hydrophobic head groups are now compressed in the middle of the bilayer
by the effective pressure. The high entropy of the lamellar phase is
taken up by the random alkyl chains, while the compressed glycerol
heads inside
adopt a two-dimensional close-packed conformation (2D hexagonal lattice).
Therefore a single wide-angle reflection at 4.17\AA corresponding to
the closest distance of
approach of glycerol heads in a plane is observed.\cite{book_Garti}
The second, nearby peak at 4.11\AA is the other characteristic
distance in this bilayer of closely packed glycerol heads, i.e. the
distance between the neighboring heads in different layers. The phase
transition from the isotropic to this inverse-lamellar phase would
still be a reversible first order phase transition.

The most dramatic difference of MG solutions in oil (in contrast to MG/water,
or most other surfactants in oil) is that the inverse lamellar phase macroscopically
behaves as an elastic gel, unlike the complex-fluid rheological behavior of
ordinary flexible lamellae. In most other surfactant solutions, and certainly
in aqueous MG systems, the gelation only occurs at lower temperatures due to
 crystallization of phases. We assert that the origin of this inverse-lamellar gelation
is in the 2D hexagonal ordering of glycerol heads inside the bilayer that sets in
from the moment the bilayer is formed. In hydrated bilayer lamellae this is not
possible, and most other surfactants that form inverse bilayers in oil have different
size ratio between the heads and the tails, and also do not achieve this 2D
crystallization on dense-packing.

The next,
lower-temperature phase transition occurs on cooling the inverse
lamellar phase below its crystallization point. The hexagonal
lateral packing of extended molten chains in the dense brush
transforms into a sub-alpha crystal form, which has
orthorhombic chain packing, characterized by strong X-ray short
spacing at 4.17\AA and several spacings from 4.06 to
3.6\AA.\cite{book_Garti} These dimensions are represented in the
Xray scattering pattern we obtain below $T_K$, in the phase that we
continue to call ``sub-alpha crystal'' to preserve the
analogy with the well-studied phase of MG in water.\cite{Larsson67, Larsson90, Sein02,
Hendricx98,Marangoni07a, Marangoni06b, Marangoni07c} In aqueous MG systems
this phase is also found, but much lower in the phase sequence, essentially
when the bilayer lamellae with crystallized tails parallel-pack so close together
that the glycerol heads on their are able to establish a 2D lattice in the contacting
planes. In our water-free system, the 2D hexagonal order sets in before the
crystallization of aliphatic tails, and so sub-alpha phase is adjacent to the
inverse lamellar phase.


In this paper different techniques were applied to provide a
comprehensive set of measurements for understanding MG/oil systems.
Firstly, in order to distinguish the MG/oil systems from water-containing
systems, we demonstrate that the presence of water, even in small
quantity (0.5\% w/w), gives a significant change of phase behavior.
We examine MG/oil systems over the whole
range of surfactant concentrations, but specifically focus on the
range 0 to 10\%, because this is the most important for region for
applications. The phase diagram was obtained by calorimetry with
high resolution. Two universal transitions, gelation and
crystallization, were indicated in different conditions (different types of
oil, varying cooling/heating rate). The crystallographic
structure in different phases was
determined by high-resolution X-ray diffraction. We find two unique ordered
states:
 the inverse lamellar phase (with hexagonal head packing) and ``sub-alpha''
crystalline state. The rheological properties in different phases
and across phase transition boundaries have been studied to give a
phenomenological description of mechanical properties. A particular feature, again
in stark contrast with water-MG solutions is that the inverse lamellar phase
rheologically behaves as a stiff gel, due to its 2D lattice of closely packed
heads inside each bilayer.

\section{Experimental details}


Distilled saturated MG were purchased from Palsgaard A/S (Denmark).
The sample contained 92\% monoglyceride C18 and 5\% C16. The
remaining 3\% consisted mainly of diglycerides and small amounts of
triglycerides. Pure MG C16 (99+\% pure) had been prepared by
chemical synthesis in our laboratory\cite{Korean01} to compare with the
commercial materials (which are always impure to some degree).
Two hydrophobic solvents were used in the study: natural
hazelnut oil, and n-tetradecane as a pure model oil. The hazelnut
oil was obtained from Provence (France) where this variety contains
approximately 80\% oleic and 20\% linoleic acids with low quantities
of MG.\cite{oil} This oil crystallizes at a temperature below
$-23^{\circ}$C. Before testing the hazelnut oil was heated to
$120^{\circ}$ for several hours to keep its drying condition.
n-tetradecane was purchased from BDH chemicals Ltd (Poole, UK) with
a quoted purity of 99\%; the freezing point of this model
hydrophobic solvent is $5.9^{\circ}$C. The mixtures of MG and oil
were stored on a heating plate at a constant temperature of
100${}^{\circ}$C, in a dessicator with a magnetic stirrer.

Later on, in the discussion of our results, one may question how
important was the small impurity of the MG in the sequence of phase
transformations. We are clear that this is not important at all. To
test this point we have separately investigated the phase
transitions of pure MG C16 in n-tetradecane. There are small
quantitative changes (reported and discussed later in the text), but
the generic feature of two consecutive phase transitions, and the
structures of ordered phases, remains universal.

Heat exchange involved in a phase transition yields exothermic or
endothermic peaks that were recorded in a differential scanning
calorimeter (DSC) experiment. From these measurements the transition
temperatures can be estimated with good resolution. A Perkin-Elmer
power-compensated Pyris 1 DSC equipped with an Intracooler 2P was
used. To focus on the interested phase transitions, samples were
heated to 100${}^{\circ}$C, held for 1 min, cooled to 0${}^{\circ}$C
at a specified rate, and then reheated to 100${}^{\circ}$C at the
same rate. The experiment was then repeated at a different
heating/cooling rate with the same sample.

The X-ray scattering patterns were recorded at different
temperatures for 10\% (w/w) monoglyceride/oil sample. Small-angle
X-ray diffraction (SAXS) was performed using a copper rotating anode
generator (Rigaku-MSC Ltd) equipped with X-ray optics by Osmic Ltd.
Before recording the X-ray diffraction, the samples were heated to
70${}^{\circ}$C for 5 min to erase the structure memory and then
cooled down to a given temperature. The distance between the small
angle detector to the sample was set to 300mm, giving the maximum
resolution of 3.36\AA at the edge of the diffraction pattern.
Samples, of thickness 1mm, were held between mica sheets of 0.1mm
thick (supplied by Goodfellow, Cambridge, UK) and an aluminium
plate. A metal substrate plate was used to ensure accurate heat
transfer to the sample. The temperature was controlled by a
home-made chamber and verified by a thermocouple. The bilayer
lamellar spacing and crystallographic lengths of crystalline phases
were calculated from the diffraction patterns using
Bragg scattering analysis.\cite{Ghosh83}

Rheological measurements were staged on a strain-controlled
rheometer (Rheometrics PHYSICA MCR501, Anton German) connected to a
water-bath temperature control, an acceptable source since our
working range was between 70${}^{\circ}$C to 20${}^{\circ}$C. A
plane-plane sensor design of annular gap 0.4mm was utilized to
ensure a constant shear rate in the total volume of the liquid. The
sample used in the rheological measurements was 10wt\% MG/oil. All
tests were carried out using fresh samples which were pre-sheared in
the rheometer with a stress of 10 Pa while kept in the isotropic
phase at 70${}^{\circ}$C for 30 min.\cite{Eugene02a,Eugene02b} When
the MG/oil were heated to an isotropic phase, all previous
orientations are removed to ensure that the samples all have the same
thermal history. The temperature ramp test involved
observing the rheological transformations at the boundary of phase
transitions. Measurements were performed under low amplitude
oscillatory shear at a low frequency of 1 rad/s with an initial
applied strain amplitude 0.05\%. These test conditions were well within
the linear viscoelastic range as determined by stress sweeps at
70${}^{\circ}$C and 26${}^{\circ}$C. Samples were steadily cooled
from 70${}^{\circ}$C to 20${}^{\circ}$C at stepped cooling rate of
1.0${}^{\circ}$C/min and the evolution of the loss and storage
moduli was monitored.

\section{Phase diagram}

\begin{figure}
\begin{center}
\resizebox{0.6\textwidth}{!}{\includegraphics{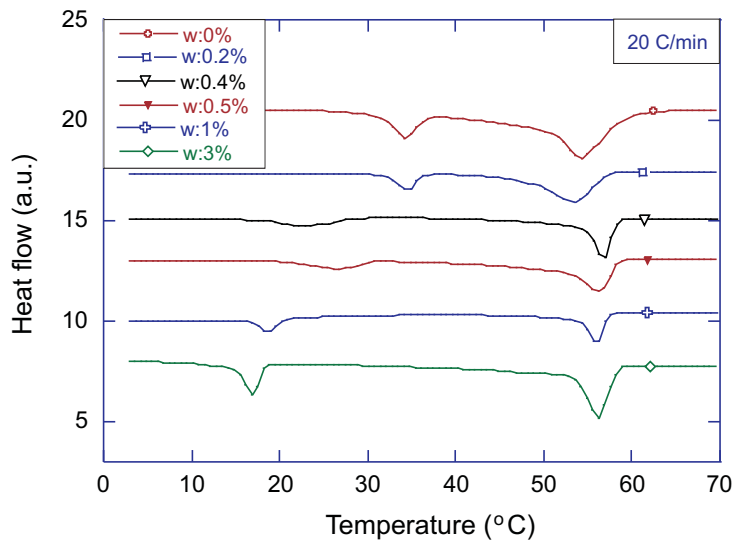}}
\caption{Above a critical concentration of water, 0.5\% (w/w) the
crystallization temperature $T_K$ decreases from 36${}^{\circ}$C to
18${}^{\circ}$C. The molar estimate shows that this is the concentration
at which there is one water molecule for each molecule of MG.} \label{fig0}
\end{center}
\end{figure}

Before the discussions of completely water-free environment (the main subject of this paper),
MG/oil mixtures with different concentrations of water wil be discussed here.
MG/water/oil systems were produced by vigorously mixing a hot oil-MG
solution with distilled water at temperature of 80${}^{\circ}$C for 20mins.
For this study, the samples always contained 10\% (w/w) MG in oil but
different concentrations of water were applied.
Calorimetric results showing the changes of the phase sequence on adding water
are shown in Fig.\ref{fig0}. Below 0.5\% (w/w) of water in the MG/oil system
there were no significant changes in the DSC cooling scans. Above the water
concentration of 0.5\% (w/w) the low-temperature transition shifted from
36${}^{\circ}$C to 18${}^{\circ}$C and the shape of the peaks became
increasingly sharp. In order to understand this data, the mol concentrations
were calculated in the following: 10wt\% MG corresponded to 0.2793M
(mol/L) and 0.5wt\% water corresponds to 0.277M (mol/L). Therefore, the
change in phase behavior occurs when there is, crudely, one water molecule
for each molecule of MG. When the ratio of water/MG molecular numbers was
above one, the changes of phase behaviors was
significant. Water has a strong affinity to the glycerol groups and, therefore,
would forming a thin layer inside the inverse lamellar ordering. We suggest that
at sufficient amount of this ``internal hydration'', the packing of glycerol heads
is disturbed, which causes the Krafft temperature shift to a lower value.
However, further detailed studies of MG/oil/water systems are needed to confirm this
hypothesis. Here the result simply highlights the fact that even small amounts of
water in MG/oil would dominate the phase behavior of mixtures. At the same time,
Fig.\ref{fig0} shows that if we keep the water content below 0.2\% (w/w), the
system may be considered water-free for all practical purposes. This is the
system we discuss from now on.

\begin{figure}
\begin{center}
\resizebox{0.6\textwidth}{!}{\includegraphics{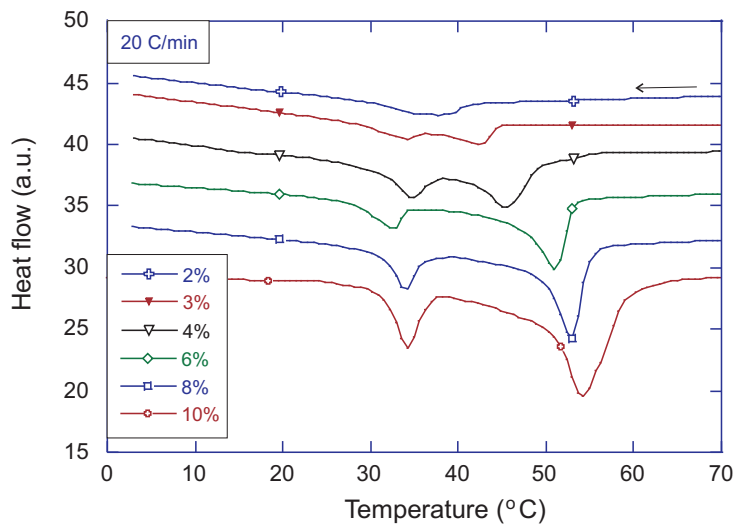}}
\caption{Collated DSC scans of MG C18 in hazelnut oil, at varying
concentrations labeled on the plot, at the cooling rate
20${}^{\circ}$C/min. There are two sequential phase changes, the
first -- of the inverse-lamellar transition and the
second -- the crystallization transition.
Below 2wt\%, the window of the molten lamellar phase
disappears as its transition dropped below the Krafft
temperature.} \label{fig1}
\end{center}
\end{figure}

\begin{figure}
\begin{center}
\resizebox{1.0\textwidth}{!}{\includegraphics{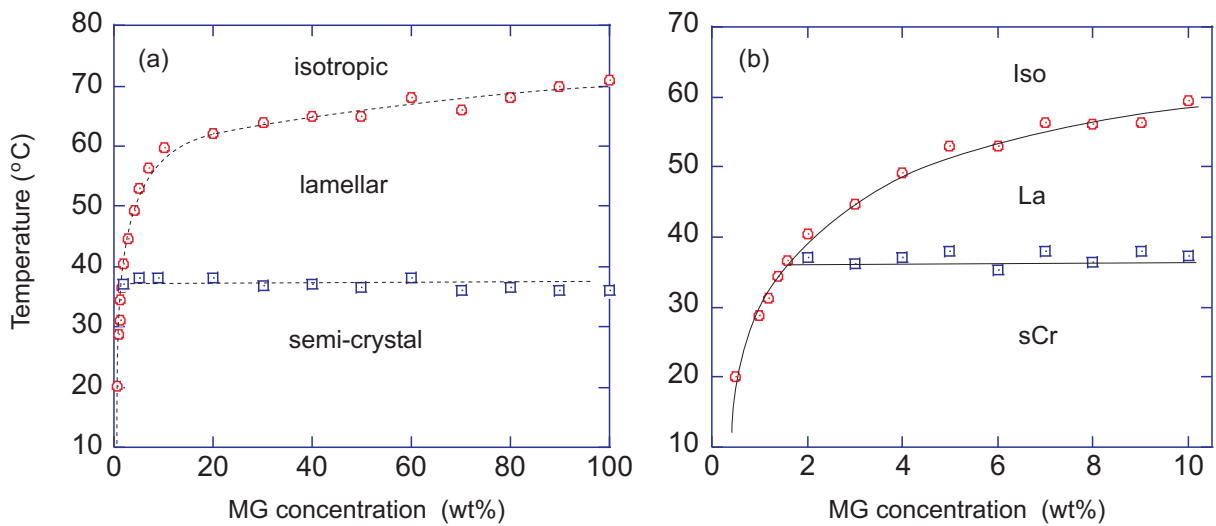}}
\caption{The phase diagram of C18 in hazelnut oil,
from the
DSC data. We show boundaries of the three phases across the
whole range of concentrations (a). A more detailed study at low
concentrations identifies the phases as: isotropic fluid, inverse
lamellar and the sub-alpha crystalline phases. Below $\sim 2$wt\%
the carbon chains of MG crystallize directly.} \label{fig2new}
\end{center}
\end{figure}

DSC experiments were performed on samples of different
concentrations of MG/oil, with the typical results collated in
Fig.\ref{fig1}. Based on this result, a phase diagram could be
sketched, Fig.\ref{fig2new}. Similar to the results found in the
MG/water solutions, two transition peaks are observed between
0${}^{\circ}$C and 100${}^{\circ}$C. In our case, the high-temperature transition
corresponds to the gelation temperature of the inverse lamellar
phase, and the second (low-temperature) transition represented the
Krafft temperature $T_K$ at which the aliphatic chains crystallize in
the lamellae. Therefore if a sample of 10\% w/w MG in oil was cooled
from 100${}^{\circ}$C, the isotropic fluid phase remained until the
temperature reached 60${}^{\circ}$C, when an inverse lamellar
ordering was formed. In this phase, due to the effective pressure from the bilayer,
the glycerol head-groups arrange with each other in a closely-packed
manner and force the alkyl chains packing to the dense-brush configuration
on the outside of the bilayer. The dense polymer
blush arrangement is important feature in this discussion: it causes
the physical properties of inverse lamellar
phase to be very different with the usual lamellar ordering. In our
case the grafting (head-group) plane of this brush has crystallographic
order and thus does not fluctuate as the fully hydrated monolayer of
glycerol head on the outside of the ordinary bilayer lamella. This, combined with
the aliphatic tails being fully extended, results in the gel-like
macroscopic properties of the inverse lamellar phase.

When the material is further cooled below
36${}^{\circ}$C, the temperature of demixing was reached. The alkyl chains
arrange in parallel sheets to form the structure analogous to what is usually
called ``sub-alpha'' crystalline phase in MG/water solutions.
By identifying the transition temperatures at different
concentrations of MG, the phase diagram of concentration-temperature
could be determined. From the phase diagram we find that
increasing the MG concentration shifts the gelation (inverse-lamellar)
transition upwards. However, the Krafft temperature for crystallizing alkyl
chains is essentially independent of the concentration of the MG.

\begin{figure}
\begin{center}
\resizebox{0.6\textwidth}{!}{\includegraphics{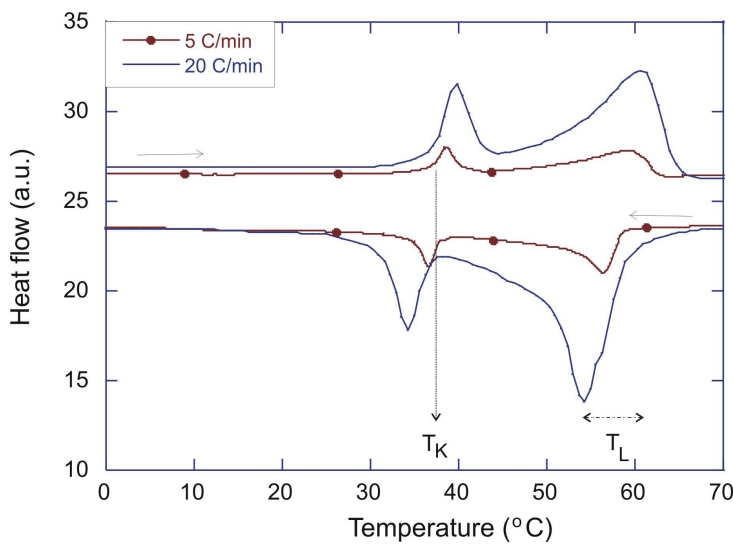}}
\caption{Heating and cooling DSC scans of 10\% (w/w) MG/oil system,
obtained at the rates of 20${}^{\circ}$C/min (solid line) and
5${}^{\circ}$C/min (circles). Since the transition points are
defined at the onset of each calorimetric peak in the respective
directions of temperature change, we conclude that hysteresis is
significant during the isotropic-lamellar transition, $T_{\rm L}$,
but is weak during the lamellar-crystallization $T_{\rm K}$. }
\label{fig3}
\end{center}
\end{figure}

Thermal hysteresis is the difference between superheating and
supercooling temperatures of a first-order phase transition,
reflecting the metastability and the height of thermodynamic barrier
between the two phases. Hysteresis of both the isotropic-lamellar and
the crystallization transitions in our system is illustrated in
Fig.\ref{fig3}. The thermal hysteresis of isotropic-lamellar
transition was significant and strongly depended on the
heating/cooling rate, increasing when the rate of heating/cooling
increased. The crystallization transition had a much weaker
hysteresis, under 1${}^{\circ}$, which practically did not
change at different heating or cooling rates that we could apply to
this system.

\begin{figure}
\begin{center}
\resizebox{0.6\textwidth}{!}{\includegraphics{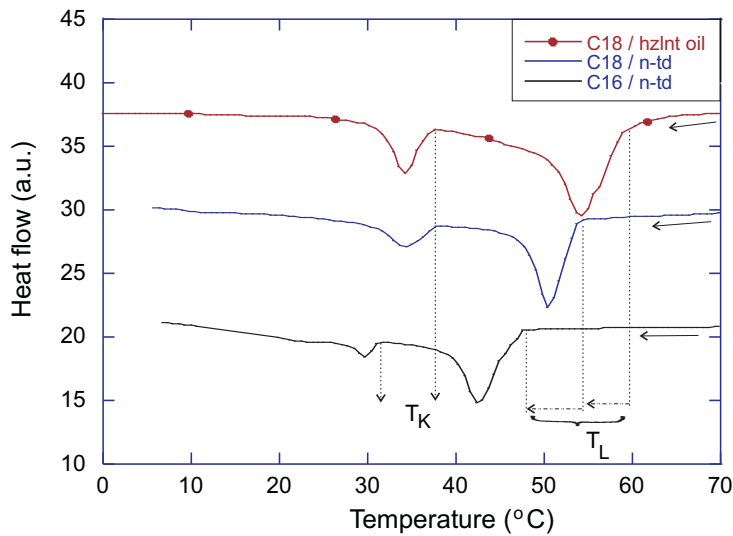}}
 \caption{Illustration of the universality of the two
transitions sequence at the cooling rate 20${}^{\circ}$C/min (the
mixtures are labeled in the plot). Clearly
the same phase sequence was observed in all systems. A small
shift of the gelation temperature $T_{\rm L}$ of C18 in the hazelnut
oil was observed; a shift to lower temperatures was found for both
transitions for smaller molecule of C16. } \label{fig4new}
\end{center}
\end{figure}

It is important to verify the universality of these findings. For
this purpose, a model oil, n-tetradecane, and pure surfactant MG C16
have also been considered to compare with the
commercial (not perfectly pure) MG C18 and hazelnut oil. Based on
many industrial applications, hazelnut oil was a natural choice for our research
target. However due to a complicated mixture of different fatty
acids in hazelnut oil, it is necessary to test a pure hydrophobic
solvent to confirm that the results from hazelnut oil are acceptably
generic for a wider range of hydrophobic solvents. The comparison of
phase transformation between 10wt\% MG/hazelnut oil and 10wt\%
MG/n-tetradecane is given in Fig.\ref{fig4new}. The transition
sequence was indeed very similar. As we expected, due to the
presence of small quantities of MG in the hazelnut oil itself, the
gelation temperature in hazelnut oil background shifted to slightly
higher values. This is consistent with the phase diagram results
above, in which a higher MG concentration would shift the gelation
temperature upwards. Also as expected, the Krafft crystallization
temperature did not change at all This
suggests that the results reported here could be used as a guide
for many other oil solvents. The transitions of pure MG C16 in
n-tetradecane also shown in Fig.\ref{fig4new} to compare with the
commercial MG C18. The result shows that the gelation temperature
to form the inverse lamellar phase decreased, as did the
Krafft temperature, by the comparable amount -- which is indeed
expected due to the shorter aliphatic tail in C16. This
evidence suggests that MG with different length of carbon chains
still retains the same features of phase transitions and confidently
concluds that two crystallization peaks observed in our
thermograms did not occur due to impurities or mixing of different
monoglycerides.

\section{Structure and rheology}

\begin{figure}
\begin{center}
\resizebox{0.8\textwidth}{!}{\includegraphics{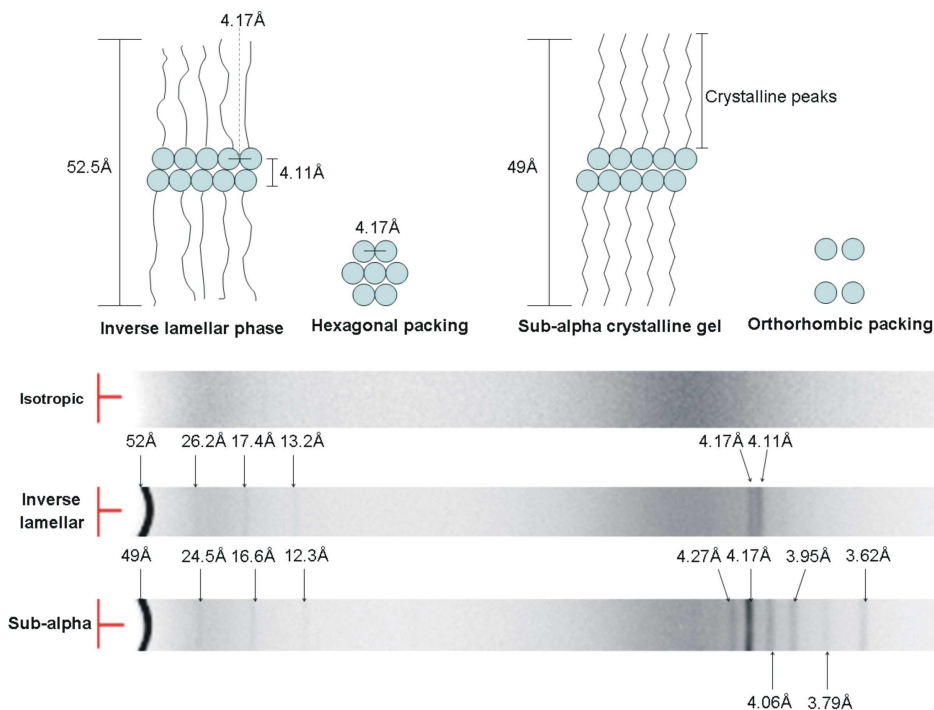}}
\caption{In the isotropic phase, there are no significant
ordering and the scattering pattern represents
just the oil background. In the inverse lamellar phase and sub-alpha
crystalline phases, the concentric rings in the small-angle region
confidently describe the lamellar ordering. In the inverse lamellar
phase the twin wide-angle peaks could be observed at 4.17\AA and
4.11\AA. They corresponded to the hexagonal packing of in-plane
glycerol heads and the ordering of plane-plane glycerol heads, which
cannot be found in the ordinary hydrated lamellar phase. The sub-alpha
crystalline phase was in orthorhombic chain packing which could be
characterized by a series of peaks between 4.27 and 3.62\AA,
with a strong wide-angle peak at 4.17\AA still reflecting the
2D dense packing of head-groups. } \label{fig5_2}
\end{center}
\end{figure}

\begin{figure}
\begin{center}
\resizebox{0.9\textwidth}{!}{\includegraphics{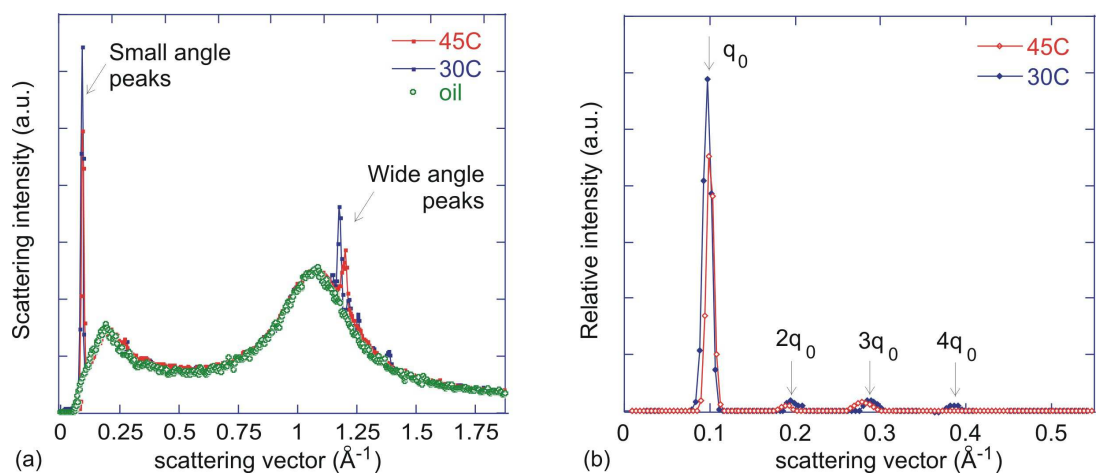}}
\caption{The cross section intensity scans of X-ray diffraction
patterns. (a) The original X-ray diffraction
includes the broad band background of the hazelnut oil solvent. (b)
By subtracting the oil background, the resulting small-angle peaks
are clear and show the high-order reflections at correct positions
reflecting the lamellar bilayer structure. The similarity of the
layered structures in the inverse lamellar phase and the sub alpha
crystal is also evident. } \label{fig6}
\end{center}
\end{figure}

The mixtures of MG in oil show diverse structuring in different
phases; these could be fingerprinted in X-ray diffraction patterns in
Fig.\ref{fig5_2} and Fig.\ref{fig6} To compare with several important studies in
the aqueous systems we collated data from various sources to highlight
 the difference between
oil-based and aqueous systems by indicating the X-ray peak positions
in the Table\ref{T1}. Our X-ray scattering results show that between the
gelation and the Krafft temperatures, MG molecules aggregated in the inverse
lamellar bilayers surrounded by oil, with
hexagonal close-packed ordering of surfactant heads in the middle plane.
Below the Krafft temperature, the inverse lamellar phase was crystallized and
transformed into a sub-alpha crystal form, which contained
orthorhombic packing of aliphatic chains.

\begin{table}[!ht]
\centering \caption{Comparison of X-ray diffraction data for MG/oil and MG/water.
\label{T1}}
\begin{tabular} {c c c}
\textbf{Systems} & \textbf{Long spacings (\AA)} & \textbf{Short spacings (\AA)} \\
\hline
MG/water(Lamellar) & 48.5\AA & -\\
MG/water(alpha crystal) & 54.3\AA & 4.18\AA\\
MG/water(beta crystal)& 48.5\AA & 4.60-4.38-4.31-4.04\AA \\
MG/oil(Inverse lamellar) & 52\AA & 4.17-4.11\AA \\
MG/oil(sub-alpha crystal) & 49\AA &  4.27-4.17-4.06-3.95-3.79-3.62\AA\\
\end{tabular}
\end{table}

Above the gelation temperature (57${}^{\circ}$C) the sample was in
an isotropic phase with no X-ray scattering features and the
powder diffraction pattern similar to the one from the
pure hazelnut oil. At 45${}^{\circ}$C, when the concentration of
MG was above 2wt\%, the inverse lamellar phase appeared. In this phase
MG heads packed together with 2D hexagonal ordering and forced the
carbon chains to extend due to dense lateral confinement,
to from several lamellar plates.
Above a 4wt\% concentration of MG, these plates formed a firm but
brittle gel network capable to hold the oil inside. Although
Fig.\ref{fig5_2} reports only on the 10wt\% sample, we have seen the
same lamellar spacing scattering in all such systems, but with much
lower intensity making it more difficult to present. A series of
concentric rings in small-angle region was sufficient to determine the
lamellar ordering. After subtracting the oil background,
Fig.\ref{fig6} we find a series of diffraction peaks with the ratio
of spacing following the sequence of 1, 1/2, 1/3, 1/4. These peaks
represented successively higher-order reflections from the periodic
lamellar structure and were in good agreement with the expected
thickness of the lamellar bilayer, determined as around 53\AA.

At 40${}^{\circ}$C, in the inverse lamellar phase, twin X-ray
diffraction peaks were observed in the wide-angle scattering region.
This is an important feature. The
 two rings indicate the regular spacings at 4.17\AA and 4.11\AA,
 which in fact characterize the spacings between neighboring glycerol heads.
 Within the same layer, the 4.17\AA spacing is characteristic of a
 2D (dense-packing) hexagonal order, while the 4.11\AA line corresponds to the
 distance between glycerol heads in the two neighboring planes inside
 the bilayer.

Below the Krafft temperature, at 35${}^{\circ}$C, the sub-alpha
crystalline phase appears. In this case the structure were still characterized by
inverted lamellar bilayers, but with the thickness of each bilayer
slightly reduced to 50\AA as indicated by small-angle scattering lines.
The orthorhombic crystallized chain packing pronounced sequence of wide
angle diffraction peaks, with a strong line at 4.17\AA and several weaker
peaks between 4.27 and 3.62\AA. Obviously this phase, which we continue to
call ``sub-alpha crystal'' is not the well
known anhydrous crystalline form, beta-crystal (with an orthogonal subcell),
which would be characterized by the sharp wide angle peaks at
4.60-4.38-4.31-4.04\AA \cite{book_Garti}.

\begin{figure}
\begin{center}
\resizebox{0.55\textwidth}{!}{\includegraphics{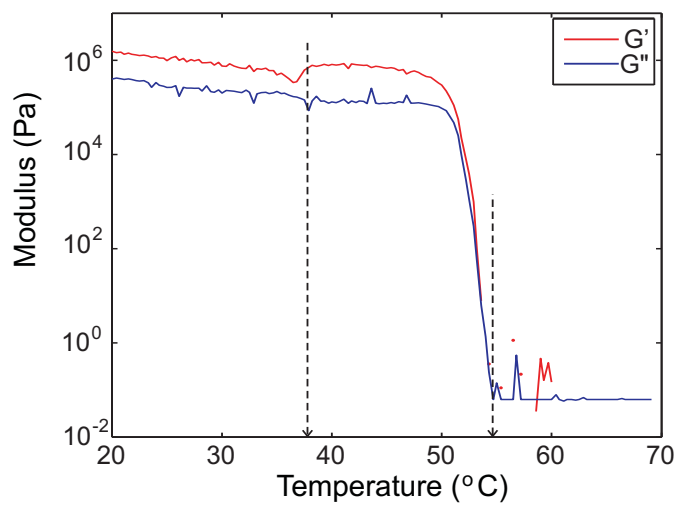}}
\caption{ There are three regions in a low-frequency rheological
behavior of 10wt\% MG/oil which correspond to the three phases.
Above 57${}^{\circ}$C the sample is in the isotropic fluid phase.
From 57${}^{\circ}$C to 38${}^{\circ}$C, the modulus increased to
$10^3$Pa while the inverse lamellar ordering consolidated. After
the crystallization temperature of 38${}^{\circ}$C was reached,
a small drop of the moduli at the phase transition is registered;
 apart from that, the rheological response of an elastic gel
was similar to the lamellar
phase.}\label{fig7}
\end{center}
\end{figure}

Rheological experiments were carried out in the linear viscoelastic
region. The measurement involved observing the low frequency shear
modulus changes as a function of temperature, as the system evolved
from isotropic to the inverse lamellar, and further to the sub alpha crystalline
phases. This provided information about macroscopic mechanical rigidity of the
systems. The materials were heated to 80${}^{\circ}$C, well into the
isotropic phase, and pre-sheared at 10Pa for 30min to erase the
thermal and mechanical history. The samples were then
cooled through two phase transition zones at a chosen well
controlled rate (1${}^{\circ}$C/min). A typical result is
illustrated in Fig.\ref{fig7}. The plot clearly shows three phases.
In the high-temperature region, the sample was in isotropic fluid
phase; the loss modulus $G"$ is higher than the storage modulus $G'$
and both are in the range of $10^-2$ Pa (we have not specifically focused
on that liquid region and the data is noisy due to the low modulus values).
When the temperature
drops down below the isotropic-lamellar transition point $T_{\rm L}$,
the rapid increase in mechanical
rigidity is immediately expressed by the storage modulus. The system
acquires mechanical characteristics of a gel. On
continuously cooling down towards the crystallization temperature
$T_{\rm c}$, a region of pre-transitional effect was observed, as
is common during nucleation at first-order transitions and
reflecting the increase in dissipation in a fluctuating system.
Below the crystallization transition the sample is in sub-alpha
crystalline phase. Because the percolating network structure has kept
its shape to encapsulate the liquid oil inside the lamellar scaffold,
the mechanical response in this
phase did not show much difference with the inverse lamellar phase.

\section{Conclusion \label{sec:5}}

To the best of our knowledge, this is the first systematical study
of the phase behavior of MG/oil systems. Calorimetric, X-ray, and
rheometric data gathered in this work yield a comprehensive set of
macroscopic and microstructural characteristics. First of all,
we identified the significant change in properties by adding water to the
MG/oil system and outlined the boundary (of water content) below which
the system can be considered non-aqueous. Therefore MG/oil solution
should be distinguished from any water-containing MG system.
The phase diagram in concentration-temperature
variables, determined by DSC, showed the three essential
MG/oil phases in the whole region of the surfactant concentrations.
These phases were separated by two lines of first order phase
transitions; the lower transition line being the Krafft
temperature, independent of surfactant concentration. The thermal
hysteresis of the two transitions was tested at different
heating/cooling rates. We found that lamellar transition had a
significant thermal hysteresis, but the Krafft temperature did not.
These conclusions were quite general, since the study of MG in two
very different oil presented very similar phase behavior. The
measurements of phase transitions of MG C16/n-tetradecane generalized
the result to different carbon-chain length of MG and clearly
suggest that the sequence of the two transitions was not arising from the
impurities of MG but based by the generic phase ordering of MG in hydrophobic fluid
matrix.

X-ray diffraction proved the existence of two phases, both characterized by
the rigid lamellar network spanning the whole volume:
the inverse lamellar, and the sub-alpha crystalline phase. The most
important finding is that in the
inverse lamellar phase the glyceride groups are densely packed in the
hexagonal manner in the planes in the middle of bilayers. This explains
the two ``twin'' wide-angle reflections corresponding to spacings of
4.17\AA and 4.11\AA, which
characterize the regular 2-dimensional packing of in-plane and plane-plane
surfactant heads. Due to the orderly hexagonal packing of glycerol
heads, the carbon chains are forced to form the dense brush ordering.
Due to this added rigidity of lamellar bilayers,
 the rheological behavior of the inverse lamellar
phase was similar with the gel-like materials known at lower temperatures in the
aqueous systems. Below the crystallization point, the lamellar phase transforms into the
sub-alpha crystalline phase, which has orthorhombic packing of aliphatic chains.
There is no significant change in the rheological response of a gel in this phase.

Apart from the detailed results and characterization of phases, the important
message of this work is that although the phase sequence of MG ordering in oil
is superficially similar to that in water, there are important differences in
the phase structure, both in the lamellar and in the low-temperature crystalline
phases. The origin of these differences lies in the inverted lamellar nature, which
(in contrast with the hydrated lamellae) does not allow sufficient fluctuations
of glycerol heads and results in much higher ordering even in the lamellar phase.
It also leads to the reversal of order between crystallization of heads and
aliphatic tails, so the first crystalline phase that appears below the lamellar
is the ``sub-alpha'' phase (which exists much lower down the phase sequence in aqueous
systems).
In this work we only studied the phases that establish immediately, on in a short time
after cooling through phase transitions. A subsequent paper will focus on the
effects of long-time aging in both phases, and the nature of the global equilibrium
order of MG in hydrophobic environment, which is controlled by intra- and
inter-glycerol hydrogen bonding of monoglyceride.

\subsection*{Acknowledgements}

\thanks{We acknowledge S.M. Clarke and A.R.
Tajbakhsh for useful discussions and guidance. The help of D.Y.
Chirgadze, in obtaining the SAXS X-ray data is gratefully
appreciated. This work has been supported by Mars U.K. }
%

\end{document}